**I - Matter, antimatter and geometry**

**II – The twin universe model : a solution to the problem of negative energy particles**

**III – The twin universe model, plus electric charges and matter-antimatter symmetry**

J.P.PETIT & G.D'AGOSTINI
Retired
Laboratoire d'Astrophysique de Marseille
Observatoire de Marseille
2 Place Le Verrier
Marseille 13

F.HENRY-COUANNIER
Centre de Physique des Particules de Marseille

**Abstract :** We introduce a new dynamical group whose coadjoint action on its momentum space takes account of matter-antimatter symmetry on pure geometrical grounds. According to this description the energy and the spin are unchanged under matter-antimatter symmetry. We recall that the antichron components of the Poincaré group, ruling relativistic motions of a mass-point particle, generate negative energy particles. The model with two twin universes, inspired by Sakharov's one, solves the stability issue. Positive and negative energy particles motions hold in two distinct folds. The model is extended to charged particles. As a result, the matter-antimatter duality holds in both universes.

# I - Matter, antimatter and geometry

## I.1 Introduction.

In 1970 [1] the French mathematician Jean-Marie Souriau introduced the concept of coadjoint action of a group on its momentum space (or 'moment map'), based on the orbit method works [2]. This makes possible to define physical observables like energy, momentum, spin as pure geometrical objects. In the following we will show how antimatter can arise on pure geometrical grounds, from dynamical groups theory.

Let us begin to recall the Souriau's method, applied to the relativistic motion of a neutral mass point.

Let us start from the Gramm matrix, linked to the Minkowski space:

$$G = \begin{bmatrix} -1 & 0 & 0 & 0 \\ 0 & -1 & 0 & 0 \\ 0 & 0 & -1 & 0 \\ 0 & 0 & 0 & +1 \end{bmatrix}$$

Introduce the four dimensional motion space-time :

$$\boldsymbol{x} = \begin{bmatrix} x \\ y \\ z \\ t \end{bmatrix}$$

The Minkowski's metric is :

$$^{t}d\boldsymbol{x}\; G\; d\boldsymbol{x} = dt^2 - dx^2 - dy^2 - dz^2$$

Write the matrix representation of the Poincaré group acting on the space-time :

$$g_p = \begin{bmatrix} L & C \\ 0 & 1 \end{bmatrix}$$

where L is a Lorentz matrix and C a space-time translation vector.

Write the element of the Lie algebra of the group, its "tangent vector" :

$$\boldsymbol{d}g_P = \begin{bmatrix} \boldsymbol{d}L & \boldsymbol{d}C \\ 0 & 0 \end{bmatrix}$$

Following Souriau, the differentiation is performed at the vicinity of the neutral element of the group (L=1 and C=0).
We write $\delta C = \gamma$ and $\delta L = G\,\omega$ where $\omega$ is an antisymmetric matrix (see Annex A).

The adjoint action acting on the Lie algebra :

$$Ag_p : \delta g_p \equiv \{\, \omega\ ,\gamma \,\} \rightarrow\ : \delta g_p' \equiv \{\, \omega', \gamma' \,\}$$

defined by :

$$\boldsymbol{d}g' = g_P \times \boldsymbol{d}g_P \times g_P^{-1}$$

is mapped by (see Annex A) :

$$\boldsymbol{w}' = {}^{t}L^{-1}\,\boldsymbol{w}\,L^{-1}$$
$$g' = -LG\boldsymbol{w}L^{-1}C + Lg$$

An element $g_P$ of the Poincaré group $G_p$ is defined by 10 parameters $\{\ \pi_i\ \}$. This number corresponds to the dimension of the group. We can write the mapping :

$$Ag_p : \{\ \delta\pi\ \} \rightarrow \{\ \delta\pi'\ \}$$

The « coadjoint » action of the group on its momentum space is defined, after Souriau, as the dual of the precedent mapping.

Introduce the scalar :

$$S=\sum_{i=1}^{n} J_i \, \boldsymbol{dp}_i = \sum_{i=1}^{n} J_i' \, \boldsymbol{dp}_i' \text{ = constant}$$

whose invariance ensures the duality relationship. The momentum is a set of 10 parameters. Their number is identical to the dimension of the group.

$J_P = \{ E , p_x , p_y , p_z , j_x , j_y , j_z , l_x , l_y , l_z \}$

The first four form the four-vector P (energy-momentum) :

$$P=\begin{bmatrix} p_x \\ p_y \\ p_z \\ E \end{bmatrix} = \begin{bmatrix} p \\ E \end{bmatrix}$$

Souriau groups the last six in order to build an antisymmetric (4,4) matrix M in the following form.

$$M = \begin{bmatrix} 0 & -l_z & l_y & f_x \\ l_z & 0 & -l_x & f_y \\ -l_y & l_x & 0 & f_z \\ -f_x & -f_y & -f_z & 0 \end{bmatrix}$$

Thus the moment writes as :

$$J=\{ M , p , E \}=\{ M , P \}$$

The invariance of the scalar product gives the relations (see annex A) :

$$M' = \left(L M \, {}^tL\right) + (C \, {}^tP \, {}^tL) - (L \, P \, {}^tC)$$
$$P' = L \, P$$

which define the coadjoint action of the Poincaré group on its momentum space.

$CoAg_p : J \equiv \{ M, P \} \rightarrow J' \equiv \{ M', P' \}$

Following Souriau, we can write the 10 elements of the momentum as :

$$J_P = \begin{bmatrix} M & -P \\ {}^tP & 0 \end{bmatrix}$$

and the coadjoint action acts as :

$J_P' = g_P \; J_P \; {}^tg_P$

Souriau introduces the following vectors l and f , respectively the spin vector and the 'passage':

$$l = \begin{bmatrix} l_x \\ l_y \\ l_z \end{bmatrix} \qquad f = \begin{bmatrix} f_x \\ f_y \\ fz \end{bmatrix}$$

The dimension of the spin vector is M $L^2$ $T^{-1}$

An ad hoc change of co-ordinates makes possible to reduce the M matrix to a single term l and the 3-momentum vector to a single component p. Then,

$$J_P = \begin{bmatrix} 0 & -l & 0 & 0 & 0 \\ l & 0 & 0 & 0 & 0 \\ 0 & 0 & 0 & 0 & -p \\ 0 & 0 & 0 & 0 & -E \\ 0 & 0 & p & E & 0 \end{bmatrix}$$

Using the geometric quantization method [3][4], Souriau showed that the spin should be quantified. This work gives a geometrical understanding of the spin as a pure geometrical quantity.

## I.2 The geometrical nature of the electric charge.

Back again to Souriau ( [1] and [5] ), who adds a fifth dimension ζ to space-time ξ :

$$\mathbf{x} = \begin{bmatrix} x \\ y \\ z \\ t \end{bmatrix} \Rightarrow X = \begin{bmatrix} \mathbf{z} \\ x \\ y \\ z \\ t \end{bmatrix}.$$

Souriau builds the following non-trivial extension of the Poincaré group.

$$\begin{bmatrix} 1 & 0 & \mathbf{f} \\ 0 & L & C \\ 0 & 0 & 1 \end{bmatrix} \times \begin{bmatrix} \mathbf{z} \\ \mathbf{x} \\ 1 \end{bmatrix} = \begin{bmatrix} \mathbf{z} + \mathbf{f} \\ L\mathbf{x} + C \\ 1 \end{bmatrix}$$

The element of the Lie Algebra can be written as :

$$\mathbf{dp} = \begin{bmatrix} 0 & 0 & \mathbf{df} \\ 0 & G\mathbf{w} & g \\ 0 & 0 & 0 \end{bmatrix}$$

ω being an antisymmetric matrix.

The adjoint action of this group, acting on the Lie algebra is defined by :

$$A_p : dp \equiv \{ df , w , g \} \rightarrow dp' \equiv \{ df' , w' , g' \}$$

with

$$dp' = p \times dp \times p$$

The extension of the Poincaré group is a 11 dimensions group, whence the new momentum owns an extra scalar q :

$$J = \{ q , E , p_x , p_y , p_z , j_x , j_y , j_z , l_x , l_y , l_z \} = \{ q , J_P \}$$

The coadjoint action, acting on momentum space, dual of the adjoint action acting on the Lie algebra, is mapped as (see Annex B):

$$q' = q$$
$$P' = L\,P$$
$$M' = L\,M\,{}^tL + ( C\,{}^tP\,{}^tL - L\,P\,{}^tC )$$

The charge q is conserved. In addition we refind the relations of the Poincaré momentum coadjoint action.

The charge q can be identified as the electric charge, considered as a new geometrical quantity.

The spin features are unchanged.

As shown by J.M.Souriau [1] when the additional dimension $\zeta$ is defined modulo $2\pi$ the charge q is quantified.

## I.3 Antimatter and geometry.

Introduce now the eight connected components matrix group.

$$\begin{bmatrix} n & 0 & n f \\ 0 & L & C \\ 0 & 0 & 1 \end{bmatrix} \times \begin{bmatrix} z \\ x \\ 1 \end{bmatrix} = \begin{bmatrix} n z + n f \\ L x + C \\ 1 \end{bmatrix} \qquad with \qquad n = \pm 1$$

Why eight ? Because Poincaré's group is a four connected component one, as the Lorentz group. The inverse matrix is :

$$\begin{bmatrix} n & 0 & -n f \\ 0 & L^{-1} & -L^{-1} C \\ 0 & 0 & 1 \end{bmatrix}$$

Here again, write the element of the Lie algebra :

$$dp = \begin{bmatrix} 0 & 0 & n\,df \\ 0 & G w & g \\ 0 & 0 & 0 \end{bmatrix}$$

The coadjoint action of this new group is the dual of the mapping :

$$A_p : \{ df , w , g \} \rightarrow \{ df' , w' , g' \}$$

with

$$dp' = p \times dp \times p^{-1}$$

We get :

$$\delta \phi' = \nu \, \delta\phi$$
$$\omega' = {}^t L^{-1} \, \omega \; L^{-1}$$
$$\gamma' = - L \, G \, \omega \, L^{-1} \, C + L \, \gamma$$

Like in the precedent section the momentum owns eleven components, including the electric charge q :

$$J = \{ q , M , p , E \}$$

Here again we use the following scalar, to build the dual mapping.

$$S = q\,df + \frac{1}{2} T_r \left( M \, w \right) + {}^t P \, G \, g$$

using the same techniques, we get :

$$q' = \nu \, q$$
$$P' = L \, P$$
$$M' = L \, M \, {}^t L + ( C \, {}^t P \, {}^t L - L \, P \, {}^t C )$$

## I.4 Conclusion :

After Souriau [5], antimatter corresponds to the inversion of a fifth dimension.
We have shown that it could be linked to the following dynamical group :

$$\begin{bmatrix} n & 0 & n f \\ 0 & L & C \\ 0 & 0 & 1 \end{bmatrix} \times \begin{bmatrix} z \\ x \\ 1 \end{bmatrix} = \begin{bmatrix} n z + n f \\ L x + C \\ 1 \end{bmatrix} \qquad with \qquad n = \pm 1$$

The associated momentum is still $J = \{ q , J_P \}$, where q is the electric charge and $J_p$ the momentum associated to the Poincaré group:

$$J_P = \begin{bmatrix} M & -P \\ {}^t P & 0 \end{bmatrix}$$

with

$$M = \begin{bmatrix} 0 & -l_z & l_y & f_x \\ l_z & 0 & -l_x & f_y \\ -l_y & l_x & 0 & f_z \\ -f_x & -f_y & -f_z & 0 \end{bmatrix}$$

But, in this new dynamical group, when ν = - 1 the element reverses the fifth dimension and the electric charge.

$$q \rightarrow -q$$
$$z \rightarrow -z$$

which corresponds to C-symmetry (charge conjugation) of a couple matter-antimatter. The spin features are unchanged. Antimatter becomes geometrized.

Given a particle with

- electric charge q
- a given energy E, assumed to be positive
- 3-momentum p
- spin s

Using the C-symmetry matrix : $\begin{bmatrix} -1 & 0 & 0 \\ 0 & 1 & 0 \\ 0 & 0 & 1 \end{bmatrix}$

and referring to the momentum laws from the coadjoint action mapping, we get an antiparticle which owns

- a reversed electric charge – q
- the same energy E
- the same 3-momentum
- unchanged spin

Antimatter owns definitively a positive energy.

Last remark : Additional extensions of the group can be performed as follows .

$$\begin{bmatrix} \boldsymbol{n} & \cdots & 0 & 0 & \boldsymbol{n} f_1 \\ \cdots & \cdots & \cdots & \cdots & \cdots \\ 0 & \cdots & \boldsymbol{n} & 0 & \boldsymbol{n} f_n \\ 0 & \cdots & 0 & L & C \\ 0 & \cdots & 0 & 0 & 1 \end{bmatrix} \times \begin{bmatrix} \boldsymbol{z}_1 \\ \cdots \\ \boldsymbol{z}_n \\ \boldsymbol{x} \\ 1 \end{bmatrix}$$

Each adds a new dimension and provides a new physical observable, which could be identified to another quantum charge. This dynamic group generalizes the concept of charge conjugation. Antimatter corresponds to the action of C-symmetry terms, corresponding to ν = - 1

$$\begin{bmatrix} -1 & \ldots & 0 & 0 & 0 & 0 \\ \ldots & \ldots & \ldots & \ldots & \ldots & \ldots \\ 0 & \ldots & -1 & 0 & 0 & 0 \\ 0 & \ldots & 0 & -1 & 0 & 0 \\ 0 & \ldots & 0 & 0 & 1 & 0 \\ 0 & \ldots & 0 & 0 & 0 & 1 \end{bmatrix}$$

The matter-antimatter symmetry goes with the inversion of quantum charges and additional dimensions:

$$q_i \to -q_i$$

$$\boldsymbol{z}_i \to -\boldsymbol{z}_i$$

## II - A geometric solution to the problem of negative energy particles

### II.1 Negative energy particles.

Consider the Gramm matrix :

$$G = \begin{bmatrix} -1 & 0 & 0 & 0 \\ 0 & -1 & 0 & 0 \\ 0 & 0 & -1 & 0 \\ 0 & 0 & 0 & +1 \end{bmatrix}$$

The matrixes of the Lorentz group obey :

$${}^{t}L\, G\, L = G$$

The following matrixes belong to the group :

$$1 = \begin{bmatrix} 1 & 0 & 0 & 0 \\ 0 & 1 & 0 & 0 \\ 0 & 0 & 1 & 0 \\ 0 & 0 & 0 & 1 \end{bmatrix} \qquad A_s = \begin{bmatrix} -1 & 0 & 0 & 0 \\ 0 & -1 & 0 & 0 \\ 0 & 0 & -1 & 0 \\ 0 & 0 & 0 & +1 \end{bmatrix}$$

$$A_t = \begin{bmatrix} 1 & 0 & 0 & 0 \\ 0 & 1 & 0 & 0 \\ 0 & 0 & 1 & 0 \\ 0 & 0 & 0 & -1 \end{bmatrix} \qquad A_{st} = \begin{bmatrix} -1 & 0 & 0 & 0 \\ 0 & -1 & 0 & 0 \\ 0 & 0 & -1 & 0 \\ 0 & 0 & 0 & -1 \end{bmatrix}$$

The first is the neutral element
The second reverses space, but not time

$$A_s \times \begin{bmatrix} x \\ y \\ z \\ t \end{bmatrix} = \begin{bmatrix} -x \\ -y \\ -z \\ t \end{bmatrix}$$

The third reverses time, but not space

$$A_t \times \begin{bmatrix} x \\ y \\ z \\ t \end{bmatrix} = \begin{bmatrix} x \\ y \\ z \\ -t \end{bmatrix}$$

The last one reverses space and time

$$A_{st} \times \begin{bmatrix} x \\ y \\ z \\ t \end{bmatrix} = \begin{bmatrix} -x \\ -y \\ -z \\ -t \end{bmatrix}$$

Notice than the Gramm matrix G = $A_s$ belongs to the group.
These matrixes each belong to one of the four connected components of the group. The first, which contains the neutral element is called the neutral component and is a sub-group, called $L_n$ .
The set $L_o$ ={ $L_n$ , $L_s$ }is made of matrixes which do not reverse time. Souriau call it the orthochron set, which is also a larger sub-group.

Let us introduce a Gramm matrix :

$$\Omega = \begin{bmatrix} \alpha & 0 & 0 & 0 \\ 0 & \alpha & 0 & 0 \\ 0 & 0 & \alpha & 0 \\ 0 & 0 & 0 & \beta \end{bmatrix} \qquad with \quad \alpha = \pm 1 \quad and \quad \beta = \pm 1$$

Notice that :

$\Omega$ ( -1 , 1 ) $L_n$ = $L_s$
$\Omega$ ( 1 , -1 ) $L_n$ = $L_t$
$\Omega$ ( - 1 , -1 ) $L_n$ = $L_{st}$

The set of matrixes which reverse time does not form a sub-group. Souriau calls it the antichron subset [1].

$L_a$ = { $L_t$ , $L_{st}$ }

The general matrix of the Lorentz group is $L = \Omega ( \alpha , \beta ) L_n$

Notice that : $L_{st}$ = - $L_n$ and $L_t$ = - $L_s$

More generally : $L_{ac}$ = - $L_o$

We can write the whole group through : $L = \mu L_o$ with $\mu = \pm 1$

In the beginning of this paper the coadjoint action of the Poincaré group (following)

$$g_p = \begin{bmatrix} L & C \\ 0 & 1 \end{bmatrix}$$

on its momentum space was derived :

$P' = L P$
$M' = L M {}^tL + ( C {}^tP {}^tL - L P {}^tC )$

With :

$$P = \begin{bmatrix} p_x \\ p_y \\ p_z \\ E \end{bmatrix}$$

We can write :

$$P' = \mu \; L_o \; P$$

From which we see that the antichron components ($\mu = -1$ , corresponding to motions with reversed time" ) reverse energy.

$$\mu = -1 \Rightarrow E' = -E$$

Particles are supposed to be peculiar movements of relativistic "mass points". If these motions are ruled by the complete Poincaré group, this one includes antichron, negative energy motions known to be problematic in quantum field theory. As shown before, antimatter owns positive energy. When a particle of matter and a particle of antimatter collide they annihilate into photons. But what would happen if two particles + E and – E could collide? They would completely annihilate resulting in a vacuum state (notice that this includes the case of negative energy photons!). The problem is that nothing could also prevent the vacuum to catastrophically decay into an infinity of possible states with zero energy resulting in a very dangerous instability.

On the other hand, presently the behavior of the universe makes difficult to resist the idea that some negative energy matter (so-called dark energy) could contribute and repel the matter, creating the observed phenomena at large red shifts.

## II.2 How to solve the problem.

Let us consider a new motion space composed by the two folds cover of a V4 manifold:

$$X = \begin{bmatrix} f \\ x \\ y \\ z \\ t \end{bmatrix} \equiv \begin{bmatrix} f \\ \mathbf{x} \end{bmatrix} \qquad with \qquad f = \pm 1$$

We give the two folds the same Gramm matrix :

$$G = \begin{bmatrix} -1 & 0 & 0 & 0 \\ 0 & -1 & 0 & 0 \\ 0 & 0 & -1 & 0 \\ 0 & 0 & 0 & 1 \end{bmatrix}$$

i.e. the same Minkowski metric :

$$ds^2 = {}^t\mathbf{dx}\, G\, \mathbf{dx} = -\, dx^2 - dy^2 - dz^2 + dt^2$$

Call $L_o$ the orthochron sub-group of the Lorentz group. We know that the whole group corresponds to :

$$L = \mu\, L_o \ \text{ with } \mu = \pm 1$$

Call C le the four-vector (space-time translations) :

$$C = \begin{bmatrix} \Delta x \\ \Delta y \\ \Delta z \\ \Delta t \end{bmatrix}$$

Introduce the following new dynamical group :

$$\begin{bmatrix} m & 0 & 0 \\ 0 & mL_o & C \\ 0 & 0 & 1 \end{bmatrix} \times \begin{bmatrix} f \\ \mathbf{x} \\ 1 \end{bmatrix} = \begin{bmatrix} m f \\ mL_o \mathbf{x} + C \\ 1 \end{bmatrix}$$

The inverse matrix is given by :

$$\begin{bmatrix} m & 0 & 0 \\ 0 & mL_o & C \\ 0 & 0 & 1 \end{bmatrix}^{-1} = \begin{bmatrix} m & 0 & 0 \\ 0 & mL_o^{-1} & -mL_o^{-1} C \\ 0 & 0 & 1 \end{bmatrix}$$

Now, let us build the coadjoint action of the group on its momentum space.

Starting from the adjoint action of the group on the Lie Algebra :

$$g \times \mathbf{d}g_{\,(g=1)} \times g^{-1} = \mathbf{d}g'_{\,(g=1)}$$

with the following elements of the Lie Algebra :

$$\mathbf{d}g = \begin{bmatrix} 0 & 0 & 0 \\ 0 & m\mathbf{d}L_o & \mathbf{d}C \\ 0 & 0 & 0 \end{bmatrix} \equiv \begin{bmatrix} 0 & 0 & 0 \\ 0 & Gw & g \\ 0 & 0 & 0 \end{bmatrix}$$

where ω being an antisymmetric matrix.

$$g \times dg_{(g=1)} \times g^{-1} = dg'_{(g=1)}$$

gives

$$\begin{bmatrix} 0 & 0 & 0 \\ 0 & Gw' & g' \\ 0 & 0 & 0 \end{bmatrix} = \begin{bmatrix} m & 0 & 0 \\ 0 & mL_o & C \\ 0 & 0 & 1 \end{bmatrix} \times \begin{bmatrix} 0 & 0 & 0 \\ 0 & Gw & g \\ 0 & 0 & 0 \end{bmatrix} \times \begin{bmatrix} m & 0 & 0 \\ 0 & mL_o^{-1} & -mL_o^{-1}C \\ 0 & 0 & 1 \end{bmatrix}$$

$$= \begin{bmatrix} 0 & 0 & 0 \\ 0 & m^2 L_o G w L_o^{-1} & m^2 L_o G w L_o^{-1} C + mL_o g \\ 0 & 0 & 0 \end{bmatrix}$$

with $\mu^2 = 1$ this is :

$$\begin{bmatrix} 0 & 0 & 0 \\ 0 & Gw' & g' \\ 0 & 0 & 0 \end{bmatrix} = \begin{bmatrix} 0 & 0 & 0 \\ 0 & L_o G w L_o^{-1} & L_o G w L_o^{-1} C + mL_o g \\ 0 & 0 & 0 \end{bmatrix}$$

which leads to the following mapping :

$$w' = {}^t L_o^{-1} \, w L_o^{-1}$$
$$g' = -G \, {}^t L_o^{-1} \, w L_o^{-1} C + mL_o g$$

As for the Poincaré Group, to deal with duality, introducing the following invariant scalar quantity:

$$S = \frac{1}{2} \; T_r \left( M w \right) + {}^t P G g = \frac{1}{2} \; T_r \left( M' w' \right) + {}^t P' G g'$$

gives the desired mapping for the coadjoint action of the group on the momentum space :

$P' = \mu \, L_o \, P$
$M' = L_o \, M \, {}^t L_o + \mu \, ( C \; {}^t P \; {}^t L_o - L_o \, P \; {}^t C )$

We know that P is the energy-impulsion following four vector :

$$P = \begin{bmatrix} p_x \\ p_y \\ p_z \\ E \end{bmatrix}$$

$\mu = -1$ (antichron elements ) gives $E' = -E$

Let us start from a « normal » motion, where the particle goes forwards in time, from past to future. A ( $\mu$ = - 1 ) element, from the antichron set reverses both space ( P-symmetry ) and time ( T-symmetry ) after :

$$\begin{bmatrix} m & 0 & 0 \\ 0 & mL_o & C \\ 0 & 0 & 1 \end{bmatrix} \times \begin{bmatrix} f \\ \mathbf{x} \\ 1 \end{bmatrix} = \begin{bmatrix} m f \\ mL_o \mathbf{x} + C \\ 1 \end{bmatrix}$$

but the "fold index" f is changed into – f .
The ( f = +1 ) fold is devoted to positive energy, orthochron movements.
The "twin fold" ( f = - 1 ) contains the negative energy particles, in reversed time. They cannot collide. They cannot interchange electromagnetic energy for these two folds contain their own null geodesic systems.

- In the first fold, supposed to be ours, positive energy photons cruise forwards in time
- In the second fold, the twin one, negative energy photons cruise in reversed time.

This is an expression, though group theory, of the well-known Sakharov'theory [6]. In 1967 the Russian introduces his twin universe model, composed by two twin universes owing reversed time arrows.

This action is identical to the coadjoint action of the (complete) Poincaré group on its momentum space.

$J_P$ = { M , p , E } = { M , P }

The momentum owns ten components (the dimension of the group) :

$J_P$ = { E , $p_x$ , $p_y$ , $p_z$ , $j_x$ , $j_y$ , $j_z$ , $l_x$ , $l_y$ , $l_z$ }

Refering to J.M.Souriau and introducing his two vectors :

The « passage » : $f = \begin{bmatrix} f_x \\ f_y \\ fz \end{bmatrix}$ and the « spin vector » : $l = \begin{bmatrix} l_x \\ l_y \\ l_z \end{bmatrix}$

The M matrix can be written as :

$$M = \begin{bmatrix} 0 & -l_z & l_y & f_x \\ l_z & 0 & -l_x & f_y \\ -l_y & l_x & 0 & f_z \\ -f_x & -f_y & -f_z & 0 \end{bmatrix}$$

A suitable change of co-ordinates makes possible to transform M into

$$J_P = \begin{bmatrix} 0 & -l & 0 & 0 & 0 \\ l & 0 & 0 & 0 & 0 \\ 0 & 0 & 0 & 0 & -p \\ 0 & 0 & 0 & 0 & -E \\ 0 & 0 & p & E & 0 \end{bmatrix}$$

From which the spin s appears. See [1].

From above the spins of « antichron twin particles » are identical to the spin of their orthochron sisters, from which our world is made of.

## II.3 Conclusion.

If the dynamic group of Poincaré rules the movement of mass-point particles, its antichron elements describes reversed time motions. As shown by Souriau in 1970 [1] such motions correspond to negative energy particles, for the T-symmetry reverses the energy E. On another hand, today's astronomical observations for large redshifts tend to include "dark energy", in order to explain the observed acceleration. A simple solution is to develop an idea initially expressed by Andréi Sakharov in 1967 [6]. The Russian suggested that two interacting twin folds could compose the Universe. It is possible to include the Poincaré group in this model if one assumes that the negative energy motions hold in a twin fold, which has been developed in this section II through a new dynamical group acting on a two folds space time, two fold cover of a manifold V4.

# III - The twin universe model, including matter-antimatter symmetry

## III.1 Introduction

Following Souriau (and Kaluza) [5] the electrically charged particles move in a five dimensional space ( t , x , y , z , ζ ), the last one being close. Antimatter corresponds to :

$$z \rightarrow -z$$

which goes with electric charge inversion :

$$q \rightarrow -q$$

In the part II of this paper we have built a dynamical group acting on a two-folds motions space. Let us now add the matter-antimatter symmetry.

## III.2 A dynamical group associated to a two-folds movement space, including matter-antimatter symmetry.

Introduce the following non-trivial extension of the group with its corresponding action on its momentum space, considered as the two-folds cover of a V5 manifold :

$$\begin{bmatrix} f' \\ z' \\ x' \\ 1 \end{bmatrix} = \begin{bmatrix} m & 0 & 0 & 0 \\ 0 & mn & 0 & mn\,f \\ 0 & 0 & mL_o & C \\ 0 & 0 & 0 & 1 \end{bmatrix} \times \begin{bmatrix} f \\ z \\ x \\ 1 \end{bmatrix} = \begin{bmatrix} m\,f \\ mn\,z + mn\,f \\ mL_o\,x + C \\ 1 \end{bmatrix}$$

with μ = ± 1 and ν = ± 1

The inverse matrix is :

$$\begin{bmatrix} m & 0 & 0 & 0 \\ 0 & mn & 0 & -mn\,f \\ 0 & 0 & mL_o^{-1} & -m\,L_o^{-1}\,C \\ 0 & 0 & 0 & 1 \end{bmatrix}$$

The Lie algebra element is :

$$\begin{bmatrix} 0 & 0 & 0 & 0 \\ 0 & 0 & 0 & mn\,d\,f \\ 0 & 0 & md L_o & dC \\ 0 & 0 & 0 & 0 \end{bmatrix} = \begin{bmatrix} 0 & 0 & 0 & 0 \\ 0 & 0 & 0 & mnd\,f \\ 0 & 0 & G\,w & g \\ 0 & 0 & 0 & 0 \end{bmatrix}$$

We form :

$$g \times dg_{\,(g=1)} \times g^{-1} = dg'_{\,(g=1)}$$

We get :

$$\begin{bmatrix} 0 & 0 & 0 & 0 \\ 0 & 0 & 0 & mn\,df' \\ 0 & 0 & Gw' & g' \\ 0 & 0 & 0 & 0 \end{bmatrix} = \begin{bmatrix} 0 & 0 & 0 & 0 \\ 0 & 0 & 0 & df \\ 0 & 0 & L_o^{-1} G w L_o & m L_o^{-1} G w C + m L_o^{-1} g \\ 0 & 0 & 0 & 0 \end{bmatrix}$$

therefore,

$\delta\phi' = \mu\,\nu\,\delta\phi$
$\omega' = {}^{t}L_o^{-1}\,\omega\,L_o^{-1}$
$\gamma' = -\mu\,L_o\,G\omega\,L_o^{-1}C + \mu\,L_o\,\gamma$

The duality, defined by the scalar

$$S = q\,df + \frac{1}{2} T_r\,(\,M\;w\,) + {}^{t}P\,G\,g$$

gives :

$q' = \mu\;\nu\;q$
$P' = \mu\,L_o\,P$
$M' = L_o\,M\,{}^{t}L_o + \mu\,(\,C\;{}^{t}P\;{}^{t}L_o - L_o\,P\;{}^{t}C\,)$

We see that a fold change ( $\mu = -1$ ) goes with a C-symmetry (antimatter).
Last remark : Additional extensions of the group can be performed as follows .

$$\begin{bmatrix} m & 0 & \ldots & 0 & 0 & 0 \\ 0 & mn & \ldots & 0 & 0 & mn\,f_1 \\ \ldots & \ldots & \ldots & \ldots & \ldots & \ldots \\ 0 & 0 & \ldots & mn & 0 & mn\,f_n \\ 0 & 0 & \ldots & 0 & mL_o & C \\ 0 & 0 & \ldots & 0 & 0 & 1 \end{bmatrix} \times \begin{bmatrix} f \\ z_1 \\ \ldots \\ z_n \\ x \\ 1 \end{bmatrix}$$

Each add a new dimension and provide a new scalar, which could be identified to another quantum charge. This dynamic group generalizes the concept of charge conjugation.

In our fold of the universe, antimatter corresponds to the action of C-symmetry terms :
with $\mu = 1$ and $\nu = -1$

$$\begin{bmatrix} -1 & \ldots & 0 & 0 & 0 & 0 \\ \ldots & \ldots & \ldots & \ldots & \ldots & \ldots \\ 0 & \ldots & -1 & 0 & 0 & 0 \\ 0 & \ldots & 0 & -1 & 0 & 0 \\ 0 & \ldots & 0 & 0 & 1 & 0 \\ 0 & \ldots & 0 & 0 & 0 & 1 \end{bmatrix}$$

In the twin universe, antimatter corresponds to $\mu = -1$ and $\nu = +1$

## III.3 Conclusion

In 1967 Andréi Sakharov introduced the concept of twin universe model [6]. In this part III we have shown that the duality matter-antimatter ( $\nu = -1$ and charge conjugation ) holds in the twin universe.

## ANNEX A : CO-ACTION OF POINCARÉ GROUP

Matrices of the Poincaré group, write as

$$g_p = \begin{bmatrix} L & C \\ 0 & 1 \end{bmatrix}$$

where C is a space-time translation vector :

$$C = \begin{bmatrix} \Delta x \\ \Delta y \\ \Delta z \\ \Delta t \end{bmatrix}$$

and L is a Lorentz matrix defined axiomatically with the metric matrix G by :

$$^{t}L\,G\,L = G \qquad \text{(A-1)}$$

We have the following simple formulas :

$$GG=1 \qquad \text{(A-2)}$$

$$LG = G\,{}^{t}L^{-1} \qquad \text{(A-3)}$$

$${}^{t}L^{-1} = GLG \qquad \text{(A-4)}$$

The inverse matrix is :

$$g_p^{-1} = \begin{bmatrix} L & C \\ 0 & 1 \end{bmatrix}^{-1} = \begin{bmatrix} L^{-1} & -L^{-1}C \\ 0 & 1 \end{bmatrix}$$

Write the element of the Lie algebra of the group, its "tangent vector" at the vicinity of the neutral element :

$$dg_p = \begin{bmatrix} dL & dC \\ 0 & 0 \end{bmatrix}$$

Let us differentiate the equation (A-1)

$${}^{t}\delta L\; G\,L + {}^{t}L\,G\;\delta L = 0$$

Introduce L = 1 and ${}^{t}G = G$

$${}^{t}\delta L\,{}^{t}G + G\,\delta L = {}^{t}(\,G\delta L\,) + G\,\delta L = 0$$

Whence $G\,\delta L$ is an antisymmetric matrix. We call it $\omega$ .
From this we get :

$\delta L = G\,\omega$

Write :

$\delta C = \gamma$

The adjoint action, $Ag_p$ , acting on the Lie algebra is :

$$Ag_p(dg_P) \equiv dg_p^{'} = g_P \times dg_P \times g_p^{-1}$$

$$dg_p^{'} = \begin{bmatrix} Gw' & g' \\ 0 & 0 \end{bmatrix} = \begin{bmatrix} L & C \\ 0 & 1 \end{bmatrix} \times \begin{bmatrix} Gw & g \\ 0 & 0 \end{bmatrix} \times \begin{bmatrix} L^{-1} & -L^{-1}C \\ 0 & 1 \end{bmatrix}$$
$$= \begin{bmatrix} LGwL^{-1} & -LGwL^{-1}C + Lg \\ 0 & 0 \end{bmatrix}$$

We can write this mapping for $Ag_p$ : $\delta g_p \equiv \{ \omega \ , \gamma \} \rightarrow \ \delta g_p' \equiv \{ \omega', \gamma' \}$ :

$$Gw' = LGwL^{-1}$$

Using (A-2) and (A-4), we get :

$$w' = {}^tL^{-1}\, w L^{-1} \qquad \text{(A-5)}$$

and

$$g' = -LGwL^{-1}C + Lg \qquad \text{(A-6)}$$

or else

$$g' = -Gw'C + Lg \qquad \text{(A-7)}$$

An element $g_P$ of the Poincaré group $G_p$ is defined by 10 scalars $\{ \pi_i \}$. This number corresponds to the dimension of the group. We can write the mapping :

$Ag_p : \{ \delta\pi \} \rightarrow \{ \delta\pi' \}$

The « coadjoint » action of the group on its momentum space is defined, after Souriau, as the dual of the precedent mapping.

Introduce the scalar :

$$S = \sum_{i=1}^{n} J_i\, dp_i = \sum_{i=1}^{n} J_i'\, dp_i' = \text{constant}$$

whose invariance ensures the duality relationship.
The momentum is a set of ten parameters. Their number is identical to the dimension of the group.
Souriau groups six of them in order to form an antisymmetric (4,4) matrix M.
The last four form the four-vector P (impulsion-energy)

$$J = \begin{bmatrix} p_x \\ p_y \\ p_z \\ E \end{bmatrix} = \begin{bmatrix} p \\ E \end{bmatrix}$$

Thus the moment is written as :

$$J = \{ M , p , E \} = \{ M , P \}$$

We can now write the scalar as :

$$S = \frac{1}{2} Tr ( M w ) + {}^tP G g$$

$T_r$ means « trace » which is the sum of the matrix diagonal elements.
The invariance of the scalar S gives :

$$S = \frac{1}{2} Tr ( M w ) + {}^tP G g = \frac{1}{2} Tr ( M' w' ) + {}^tP' G g' \qquad \text{(A-8)}$$

Using the precedent mapping relations (A-5) and (A-7) :

$$S = \frac{1}{2} Tr ( M' w' ) + {}^tP' G g' = \frac{1}{2} Tr ( M' \, {}^tL^{-1} w L^{-1} ) + {}^tP' G ( - G w' C + L g )$$

$$= \frac{1}{2} Tr ( M' \, {}^tL^{-1} w L^{-1} ) - {}^tP' G G w' C + {}^tP' G L g$$

with GG = 1 , we get :

$$S = \frac{1}{2} Tr ( M' w' ) + {}^tP' G g' = \frac{1}{2} Tr ( M' \, {}^tL^{-1} w L^{-1} ) - {}^tP' w' C + {}^tP' G L g$$

Identifying the $\gamma$ ( $\omega$' does not depend of $\gamma$ ) :

$${}^tP G g = {}^tP' G L g$$

$\Rightarrow$

$${}^tP G = {}^tP' G L$$

$\Rightarrow$

$${}^t({}^tP G) = G P = {}^t({}^tP' G L) = {}^t(G L) P' = G L^{-1} P'$$

$\Rightarrow$

$$P' = (G L^{-1})^{-1} G P = L G G P = L P$$

Thus :

P' = L P (A-9)

We can recognize in this formula, the mass conservation :

$$P'^2 \equiv {}^tP'\,G\,P' = {}^tP\,{}^tL\,G\,L\,P = {}^tP\,G\,P \equiv P^2$$

Consider the remaining terms in (A-8) and using (A-9), we get :

$$\frac{1}{2}Tr\left(M\ w\right) = \frac{1}{2}\ Tr\left(M'\ w'\ \right) - {}^tP'\,w'\,C = \frac{1}{2}\ Tr\left(M'\ w'\ \right) - {}^tP\,{}^tL\,w'\,C \qquad \text{(A-10)}$$

The second term of the second member of this equation is the produce of a line-matrix by a column-matrix, which can be written as :

$${}^tP\ {}^tL\,w'\,C = Tr\left(w'\,C\ {}^tP\ {}^tL\right)$$

In the trace we can operate a circular permutation :

$$Tr\left(w'\,C\ {}^tP\ {}^tL\right) = Tr\left(C\ {}^tP\ {}^tL\,w'\right)$$

(A-10) becomes :

$$\frac{1}{2}Tr\left(M\ w\right) = \frac{1}{2}\ Tr\left(M'\ w'\ \right) - Tr\left(w'\ C\ {}^tP\,{}^tL\right) = \frac{1}{2}\ Tr\left(M'\ w'\ \right) - Tr\left(C\ {}^tP\,{}^tL\,w'\right)$$

Using (A-5) we get :

$$\frac{1}{2}Tr\left(M\ w\right) = \frac{1}{2}\ Tr\left(M'\ {}^tL^{-1}\,w\,L^{-1}\ \right) - Tr\left(C\ {}^tP\,{}^tL\,{}^tL^{-1}\,w\,L^{-1}\right)$$

$\Rightarrow$

$$\frac{1}{2}Tr\left(M\ w\right) = \frac{1}{2}\ Tr\left(L^{-1}M'\ {}^tL^{-1}\,w\ \right) - Tr\left(L^{-1}\,C\,{}^tP\ w\right)$$

ω is an antisymmetric matrix. We can identify the antysymmetric factor in front of ω .
As M and M' are antisymmetric, we can see than $L^{-1}$ M' ${}^tL^{-1}$ is antisymmetric too, so we only need to anti-symmetrize the last factor.

$$antisym\left(L^{-1}C\,{}^tP\right) = \frac{1}{2}\left[L^{-1}C\,{}^tP - {}^t\left(L^{-1}C\,{}^tP\right)\right] = \frac{1}{2}\left[L^{-1}C\,{}^tP - P\,{}^tC\,{}^tL^{-1}\right]$$

then we get :

$$M = \left(L^{-1}M'\ {}^tL^{-1}\right) - \left(L^{-1}C\,{}^tP\right) + \left(P\,{}^tC\,{}^tL^{-1}\right) \qquad \text{(A-11)}$$

This formula is related to the conservation of the global kinetic moment (spin + orbital).

Finally, we get the mapping of the coadjoint action of the group on its momentum space :

| CoAg$_p$ : J ≡ { M, P } → J' ≡ { M', P' } |
|---|
| $M' = \left(L\,M\,{}^tL\right) + \left(C\,{}^tP\,{}^tL\right) - \left(L\,P\,{}^tC\right)$ |
| P' = L P |

(A-12)

Let us express the co-action CoA in such a way that the action nature is explicit.

Let us get back to the Poincaré matrix :

$$g_P = \begin{bmatrix} L & C \\ 0 & 1 \end{bmatrix}$$

Its transposed is :

$${}^t g_P = \begin{bmatrix} {}^tL & 0 \\ {}^tC & 1 \end{bmatrix}$$

Consider the following antisymmetric matrix :

$$J_P = \begin{bmatrix} M & -P \\ {}^tP & 0 \end{bmatrix}$$

Form :

$$g_P \ J_P \ {}^t g = \begin{bmatrix} L & C \\ 0 & 1 \end{bmatrix} \times \begin{bmatrix} M & -P \\ {}^tP & 0 \end{bmatrix} \times \begin{bmatrix} L & 0 \\ {}^tC & 1 \end{bmatrix}$$

We get :

$$g_P \ J_P \ {}^t g_P = \begin{bmatrix} L\,M\ {}^tL + C\ {}^tP\ {}^tL - L\,P\ {}^tC & -L\ P \\ {}^tP\ {}^tL & 0 \end{bmatrix}$$

Which can be identified to the matrix :

$$J_P' = \begin{bmatrix} M' & -P' \\ {}^tP' & 0 \end{bmatrix}$$

Therefore, the co-adjoint action of the Poincaré group can be put in a matrix form :

$$J_P' = g_P \ J_P \ {}^t g_P$$

where it is now evident to see that this is a true action of the Poincaré group.

## ANNEX B : CO-ACTION OF EXTENDED POINCARÉ GROUP

Let us add a fifth dimension ζ to space-time ξ.

$$\boldsymbol{x} = \begin{bmatrix} x \\ y \\ z \\ t \end{bmatrix} \Rightarrow X = \begin{bmatrix} \boldsymbol{z} \\ x \\ y \\ z \\ t \end{bmatrix}$$

We can build the following non-trivial extension of the Poincaré group :

$$\begin{bmatrix} 1 & 0 & \boldsymbol{f} \\ 0 & L & C \\ 0 & 0 & 1 \end{bmatrix} \times \begin{bmatrix} \boldsymbol{z} \\ \boldsymbol{x} \\ 1 \end{bmatrix} = \begin{bmatrix} \boldsymbol{z} + \boldsymbol{f} \\ L\boldsymbol{x} + C \\ 1 \end{bmatrix}$$

Whose inverse matrix is :

$$\begin{bmatrix} 1 & 0 & -\boldsymbol{f} \\ 0 & L^{-1} & -L^{-1}C \\ 0 & 0 & 1 \end{bmatrix}$$

The element of the Lie Algebra is :

$$\begin{bmatrix} 0 & 0 & d\boldsymbol{f} \\ 0 & dL & dC \\ 0 & 0 & 0 \end{bmatrix}$$

Here again, lets write :

$G\,\delta L = \omega$ and $\delta C = \gamma$

ω being an antisymmetric matrix.
Whence :

$$\boldsymbol{dp} = \begin{bmatrix} 0 & 0 & \boldsymbol{df} \\ 0 & G\boldsymbol{w} & \boldsymbol{g} \\ 0 & 0 & 0 \end{bmatrix}$$

As for the Poincaré group (cf Annex A), in order to build the coadjoint action, we write :

$$A_p : \boldsymbol{dp} \equiv \{ \boldsymbol{df}, \boldsymbol{w}, \boldsymbol{g} \} \rightarrow \boldsymbol{dp}' \equiv \{ \boldsymbol{df}', \boldsymbol{w}', \boldsymbol{g}' \}$$

with

$$dp' = p \times dp \times p$$

i.e. :

$$\begin{bmatrix} 0 & 0 & df' \\ 0 & Gw' & g' \\ 0 & 0 & 0 \end{bmatrix} = \begin{bmatrix} 1 & 0 & f \\ 0 & L & C \\ 0 & 0 & 1 \end{bmatrix} \times \begin{bmatrix} 0 & 0 & df \\ 0 & Gw & g \\ 0 & 0 & 0 \end{bmatrix} \times \begin{bmatrix} 1 & 0 & -f \\ 0 & L^{-1} & -L^{-1}C \\ 0 & 0 & 1 \end{bmatrix}$$

$$= \begin{bmatrix} 0 & 0 & df \\ 0 & LGwL^{-1} & -LGwL^{-1}C + Lg \\ 0 & 0 & 0 \end{bmatrix}$$

This gives the following mapping :

$\delta\phi' = \delta\phi$
$\omega' = {}^{t}L^{-1}\,\omega\,L^{-1}$
$\gamma' = -L\,G\,\omega\,L^{-1}\,C + L\,\gamma$

The non trivial extension of the Poincaré group is a 11 dimensional group, whence the new momentum owns an extra parameter q :

$J = \{ q , M , p , E \}$
$J = \{ q , E , p_x , p_y , p_z , j_x , j_y , j_z , l_x , l_y , l_z \} = \{ q , J_P \}$

Introducing the scalar :

$$S = q\,df + \frac{1}{2}Tr\left( M\,w \right) + {}^{t}P\,G\,g$$

Using the same calculation techniques, we get :

$$q\,df + \frac{1}{2}Tr(Mw) + {}^{t}PGg = q'df' + \frac{1}{2}Tr\left(M'w'\right) + {}^{t}P'Gg'$$

The identification term to term gives the coadjoint action mapping :

$q' = q$
$P' = L\,P$
$M' = L\,M\,{}^{t}L + ( C\,{}^{t}P\,{}^{t}L - L\,P\,{}^{t}C )$